\documentstyle[amstex,amssymb,twoside,leqno]{article}

\def\psi{\psi}

\input{tcilatex}

\begin{document}




\centerline{\bf FRACTIONAL DERIVATIVE ANALYSIS} 
\centerline{\bf OF HELMHOLTZ
AND PARAXIAL-WAVE  EQUATIONS}

\centerline{ A. J. T U R S K I ,   B. A T A M A N I U K  and  E. T U R S K A}


\centerline{Depatment of Theory of Continuous Mechanics} %
\centerline{Institute of Fundamental Technological Research, PAS}




\bigskip 

{\small Fundamental rules and definitions of Fractional Differintegrals are
outlined. Factorizing 1-D and 2-D Helmholtz equations \ four
semi-differential eigenfunctions are determined. The functions exhibit
incident and reflected plane waves as well as diffracted incident and
reflected waves on the half-plane edge. They allow to construct the
Sommerfeld half-plane diffraction solutions. Parabolic-Wave Equation (PWE,
Leontovich-Fock) for paraxial propagation is factorized and differential
fractional solutions of Fresnel-integral type are determined. We arrived at
two solutions, which are the mothers of known and new solutions.}

{\normalsize 
}

\section{{{{{{\protect\normalsize {\bf Introduction  \protect
\setcounter{equation}{0}}}}}}}}

\hspace{1.5em}The mathematical theory of the fractional calculus and the
theory of ordinary fractional differential equations is well developed and
there is a vast literature on the subject [1], [2], [3], [4], [5] and [6].
The theory of partial fractional differential equations is a recently
investigated problem and the theory mainly concerns fractional
diffusion-wave equations [7], [8], [9], [10] and [11].

The main objectives of this paper is a factorization of the Helmholtz
equation to obtain four semidifferential eigenfunctions allowing to
construct the well known half-plain diffraction problem. Factorizing the
Leontovich-Fock equation, we determine semidifferential Green functions,
which allow to find paraxial solutions for a given beam boundary conditions.

The article is organized as follows. In Sec.2, we quote the required rules
for fractional differintegrals and four fundamental definitions of
fractionals. Section 3 is devoted to{\normalsize \ }a factorization of
ordinary differential equations and determinations of fractional
eigenfunctions. Section 4 and 5 constitute the main body of our paper and
contain derivations of fractional eigenmodes for the Helmholtz equation and
fractional solutions of Leontovich-Fock equation. The final section is
devoted to comments and conclusions.{\normalsize \ }

\section{{{{{{\protect\normalsize {\bf Main Rules and Definitions of
Fractional Differintegrals\protect\setcounter{equation}{0} }}}}}}}

{\normalsize \hspace{1.5em}}Integration and differentiation to an arbitrary
order named fractional calculus has a long history, see[1], [3]. In the
background of the fractional operations, we see generalization of integer
order calculus to a noninteger order, class of differintegrable functions
and applications of the calculus. We recall required rules of fractional
differintegral calculus. For simplicity, at present let us assume that the
considered functions are real and differintegrals are of real order and are
related ti the interval of real axis from $c$ to $\infty .$

1. {\sc Analyticity: }If functions $f:R\longrightarrow R$ \ is analytic for $%
x\in (c,\infty )$ then \ the fractional of $q$-order prescribed to the
interval $(c,\infty )$ ; \ \ $_{c}D_{z}^{q}(f(x))$ is an analytic function
of $x$ and $q$.

2. {\sc Consistency: }Operation{\sc \ }$_{c}D_{x}^{q}(f(x))$ must be
consistent with the integer order differentiations if $q=n$ \ and with
integer order integrations if $q=-n$. The operation must vanish together
with n-1 derivatives at point $x=c.$

3. {\sc Zero Operation: }$_{c}D_{x}^{0}(f(x))=f(x)$

4. {\sc Linearity: }$_{c}D_{x}^{q}(af(x)+bg(x))=a\,_{c}D_{x}^{q}(f(x))+b%
\,_{c}D_{x}^{q}(g(x))$ \ where $a$ and $b$ are arbitrary constants.

5. {\sc Additivity and Translation: }$%
_{c}D_{x}^{q}(_{c}D_{x}^{p}(f(x)))=_{c}D^{q+p}(f(x)).$

The definition of fractional differintegrals due to Gr\"{u}nwald and
Letnikov is the most fundamental in that it involves the fewest restrictions
on the functions to which it is applied and avoids the explicit use of
notations of the ordinary derivative and integral. The definition unifies
two notions, which are usually presented separately in classical analysis;
the derivative and integral (differintegral). It is valid for a $q$-th order
derivative or (-$q$)-th folded integrals whether or not $q$ is a real
integer.{\normalsize \ }Thus the differintegral of $f:R\longrightarrow R$ \
is:

\begin{equation}
{\normalsize _{c}D_{x}^{q}(f(x))=\lim_{N\rightarrow \infty }\frac{h^{-q}}{%
\Gamma (-q)}\sum\limits_{j=0}^{N-1}\frac{\Gamma (j-q)}{\Gamma (j+1)}f(x-jh)%
\mbox{,}}  \tag{(2.1)}
\end{equation}

where $h=(x-c)/h,\;$and $x\geq c.$ The definition stems from the difference
quotient defining $n$-th order derivative, which contains n terms in
nominator and $n$-th power of $h\;$in denominator. The{\normalsize \ }number
of terms tends to infinity for noninteger $q$ in the nominator and $q$-th
power of $h$ in the denominator. Just like in the case of binomial formula
for the positive integer power $(a+b)^{n}\;$and for negative power as well
as noninteger power $(a+b)^{q}$. The convergence is a critical point but the
formula (2.1) is very convenient for computations. The definition is
equivalent to the Riemann-Liouville fractional integral:

\begin{eqnarray}
(I^{q}f)_{c}(x) &\doteq &_{c}D_{x}^{-q}(f(x))=\frac{1}{\Gamma (q)}%
\int_{c}^{x}(x-t)^{q-1}f(t)dt\mbox{,}  \TCItag{(2.2)} \\
0 &<&q<1  \nonumber
\end{eqnarray}%
and fractional derivative:

\begin{eqnarray}
_{c}D_{x}^{q}(f(x)) &=&\frac{1}{\Gamma (1-q)}\frac{d}{dx}%
\int_{c}^{x}(x-t)^{-q}f(t)dt\mbox{.}  \TCItag{(2.2)} \\
0 &<&q<1.  \nonumber
\end{eqnarray}

The definitions stem from consideration of Cauchy's integral formula and are
very convenient to implement but require observation of convergence of the
integrals. Nevertheless, because of their convenient formulations in terms
of a single integrations they enjoy great popularity as working definitions.
They play an important role in the development of the theory of
differintegrals and for their applications in pure mathematics-solutions of
noninteger order differential equations, definitions of new function
classes, summation of series, etc. In this paper, we do apply the these
definitions\ in spite of some ambiguity in respect to consistency and
additivity rules. However, the demands of modern applications require
definitions of fractional derivatives allowing the utilization of physically
interpretable initial conditions, which contain classical initial conditions
for functions and their integer derivatives at the initial point.
Unfortunately, the Riemann-Liouville approach leads to initial conditions
containing fractional derivatives at the lower limit of integrals. A certain
solution to this conflict was proposed by {\sc M. Caputo }[4]{\sc , }whose
definition is as follows:

\begin{eqnarray}
\left( D^{q}f\right) _{C}^{x} &=&\frac{1}{\Gamma (1-q)}\int_{0}^{x}(t)^{-q}%
\frac{d}{dx}f(x-t)dt\mbox{,}  \TCItag{(2.3)} \\
0 &<&q<1.  \nonumber
\end{eqnarray}%
where $C$ denotes the Caputo definition.

For the purpose of applications to fractional differential equations, we
introduce the Miller-Ross sequential fractional derivatives. The main idea
is based on the relation:

\begin{eqnarray*}
&&D^{n\alpha }f(x)\overset{\vartriangle }{=}\underset{n}{\underbrace{%
D^{\alpha }D^{\alpha }...D^{\alpha }}}f(x) \\
&&\text{or \ }D^{\alpha }f(x)\overset{\vartriangle }{=}D^{\alpha
_{1}}D^{\alpha _{2}}...D^{\alpha _{n}}f(x) \\
\text{where }\alpha &=&\alpha _{1}+\alpha _{2}+...+\alpha _{n}
\end{eqnarray*}%
and the simplest fractional equation of order $O(N,q)$ \ takes the form:

\begin{equation}
\sum_{j=1}^{N}a_{j}D^{jv}y(x)=f(x,y)\;\;\mbox{,}  \tag{(2.4)}
\end{equation}%
where $v=\frac{1}{q}$and the adequate initial conditions for fractional
derivatives, see [3] and [4].

We shall call (2.4) the fractional linear differential equation with
constant coefficients of the order $(N,q)$, where $q$ is the least common
multiple of the denominators of the nonzero $\alpha _{j}=jv$. The solution
to the equation can be found by use of Laplace transformations. We know that 
$N$-th order linear differential equation has $N$ linearly independent
solutions. In [3], it is shown how to construct linearly independent $N$%
-solutions of homogeneous fractional differential equations.

\section{Factorization and Eigenfunctions of an ODE{{{{{\protect\normalsize 
{\bf \protect\setcounter{equation}{0}}}}}}}}

An eigenfunction $y(\xi )$ of the linear operator $L[y(\xi )]$ is such a
function that the repeated operations preserve the function, e.g. $L[y(\xi
)]=Cy$ \ with the exactness to a multiplicative constant $C$. In the case of
fractional operation, the definition is extended to\ preservation of the
function but an additive constant or a term of power of $\xi $, e.g. $(\pi
\xi )^{-1/2}$, is subtracted at each step, see [1] and [4].

\noindent Consider the following ODE:

\begin{equation}
_{0}D_{\xi }^{2}y(\xi )=y(\xi )\mbox{,}  \tag{(3.1)}
\end{equation}%
which possesses two eigenfunction $y_{0}=e^{\pm \xi }$ and two eigenvalues $%
\pm 1$ . Factorizing the last equation according to:

\begin{eqnarray*}
(a^{4}-1) &=&(a^{2}+1)(a^{2}-1), \\
a^{2}-1 &=&(a+1)(a-1)
\end{eqnarray*}%
and 
\[
a^{2}+1=(a+i)(a-i) 
\]%
where $a=D^{1/2}$, the solution to the semi-differential equation, takes the
form:

\begin{eqnarray}
y(\xi ) &=&y_{0}(\xi )+D^{1/2}y_{0}(\xi )\mbox{,}  \TCItag{(3.2)} \\
y_{0}(\xi ) &=&e^{\xi }  \nonumber
\end{eqnarray}%
and $y(\xi )$ satisfies the semi-differential equation; $D^{1/2}y-y=0\;$as
well as the equation; $D^{2}y-y=0$ \ according to the above mentioned rules,
that is consistency and additivity. On the other hand, for $y_{0}=e^{-\xi }$
, \ we have the fractional eigenmode: 
\[
y(\xi )=y_{0}(\xi )+iD^{1/2}y_{0}(\xi ), 
\]%
which satisfies the equations; $D^{1/2}y+iy=0$ and $D^{2}y+y=0$. According
to the Riemann-Liouville definition the semi-derivative on the interval $%
x\in (0,\infty )$ is given by the formula;

\[
_{0}D_{\xi }^{1/2}e^{\xi }=\frac{1}{\sqrt{\xi }}E_{1,1/2}(\xi )=\frac{1}{%
\sqrt{\pi \xi }}+e^{\xi }Erf(\sqrt{\xi }), 
\]%
where $E_{1,1/2}(\xi )$ is a Mittag-Leffler function, see [4]. The
eigenfunction of the equation ($_{0}D_{\xi }^{1/2}y-y=0$) takes the form;

\begin{equation}
y(\xi )=e^{\xi }+\frac{1}{\sqrt{\xi }}E_{1,1/2}(\xi )=\frac{1}{\sqrt{\pi \xi 
}}+e^{\xi }Erfc(-\sqrt{\xi })\mbox{,}  \tag{(3.3)}
\end{equation}%
where the complementary error function is:

\[
Erfc(-\sqrt{\xi })=\frac{2}{\sqrt{\pi }}\int\limits_{-\sqrt{\xi }}^{\infty
}\exp (-t^{2})dt\mbox{.} 
\]%
This is a well known solution, see [4], and it is an eigenfunction of the
operator (3,1) in the sense of the extended definition of fractional
eigenfunctions. The problem is that by substitution of (3.3), we have:%
\begin{eqnarray*}
(\partial _{\xi }-1)y(\xi ) &=&-\frac{1}{2\sqrt{\pi }\xi ^{3/2}}\mbox{,} \\
(\partial _{\xi \xi }-1)y(\xi ) &=&\frac{3-2\xi }{4\sqrt{\pi }\xi ^{5/2}}%
\mbox{.}
\end{eqnarray*}%
In the next Section, we discuss the deficiency and remove power terms in the
case of the 2-D Helmholtz equation.

\section{Factorization and Eigenfunctions of Helmholtz Equations{{{{%
{\protect\normalsize {\bf \protect\setcounter{equation}{0}}}}}}}}

Let us consider the following 1-D Helmholtz equation:

\begin{equation}
_{0}D_{x}^{2}y(x)+k^{2}y(x)=0\mbox{.}  \tag{(4.1)}
\end{equation}%
Taking $x=ik\xi $ we are constructing 4-eigenfunctions by use of Fresnel
integrals $\int\limits_{x}^{\infty }exp(it^{2})dt$ instead of the
complementary error functions. We note, that in principle, there are the
following ordinary modes; $e^{ikx},\,e^{-ikx}$ and fractional modes:%
\begin{equation}
e^{ikx}\int\limits_{\sqrt{kx}}^{\infty }e^{-it^{2}}dt,\;e^{-ikx}\int\limits_{%
\sqrt{kx}}^{\infty }e^{it^{2}}dt.  \tag{(4.2)}
\end{equation}%
The first and the second pairs of modes are complex conjugate and since
there are four resulting modes: 
\[
\sin (kx),\;\cos (kx),\;\int\limits_{\sqrt{kx}}^{\infty }\cos
(kx-t^{2})dt,\;\int\limits_{\sqrt{kx}}^{\infty }\sin (kx-t^{2})dt. 
\]%
Substituting the two last terms into (4,1), the following terms; $-k^{2}/2%
\sqrt{kx}\;,\;\sqrt{kx}/4x^{2}\;$ appear on the right hand side of the
equation, respectively. The power terms are the consequences of the accepted
definition of fractal derivatives and in the considered case the consistency
and additivity rules are not completely preserved. We withhold a discussion
of any interpretation of the derived fractional modes. But for
two-dimensional problems we demonstrate a method to remove the ambiguity as
the problems have a pronounce physical meaning. In the case of the 2-D
Helmholtz equation: 
\begin{equation}
\Delta _{x,y}\Phi (x,y)+k^{2}\Phi (x,y)=0\mbox{,}  \tag{(4.3)}
\end{equation}%
it is found, that

\begin{equation}
\Phi (x,y)=e^{-iax-iby}\int\limits_{u(x,y)}^{\infty }e^{it^{2}}dt\mbox{,} 
\tag{(4.4)}
\end{equation}%
with $k=\sqrt{a^{2}+b^{2}},$ satisfies (4.3) if the function $u(x,y)$ obeys
the following characteristic equations:

\begin{equation}
u[(\partial _{x}u)^{2}+(\partial _{y}u)^{2}]=a\partial _{x}u+b\partial
_{y}u,\,\text{and \ }\partial _{x,x}u+\partial _{y,y}u=0\mbox{.}  \tag{(4.5)}
\end{equation}%
Introducing an analytical function; $f(z)=u(x,y)+iv(x,y)\;$with possible
singularity at $z=x+iy=0,$ we note, that (4.5) is the real part of the
following equation: 
\[
f^{\ast ^{\prime }}(z)f^{\prime }(z)f(z)=\kappa f^{\ast ^{\prime }}(z), 
\]%
where \ $\kappa =a+ib,$ $\left| \kappa \right| =k=\sqrt{a^{2}+b^{2}}.$ The
equation can be reduced to, called by us, an eikonal equation of diffraction;

\begin{equation}
\frac{d}{dz}(f(z))^{2}=2\kappa \mbox{.}  \tag{(4.6)}
\end{equation}%
Neglecting the constants of integration, the solution of the last equation
is 
\[
f(z)=\pm \sqrt{2\kappa z} 
\]

\begin{equation}
\text{and \ }u(x,y)=\pm \sqrt{kr+ax+by}=\pm \sqrt{2kr}\cos (\frac{\theta
-\alpha }{2})\mbox{,}  \tag{(4.7)}
\end{equation}%
where $r^{2}=x^{2}+y^{2}$, $(r,\theta )$ are polar coordinates and $\alpha $
is the incident angle of a plane wave, see [12]. In the case of reflected
waves, we have: 
\[
e^{-iax+iby}F(u_{1}). 
\]%
By use of the eikonal equation of diffraction, we obtained:

\[
u_{1}(x,y)=\sqrt{kr+ax-by}=\sqrt{2kr}\cos (\frac{\theta +\alpha }{2}). 
\]%
It deserve notice, that the obtained complex solution for the 2-D equation; $%
f(z)=\pm \sqrt{2\kappa z}\;$is analogous to 1-D case, where we have $\pm 
\sqrt{kx}$ for the lower limit of Fresnel integrals. Hence, we have
4-eigenmodes of the Helmholtz equations:

\[
e^{-iax-iby},\;e^{-iax+iby} 
\]

\[
e^{-iax-iby}F(u(x,y))\;\text{and}\;e^{-iax+iby}F(u_{1}(x,y)), 
\]%
where \ $F(u)=\int_{u}^{\infty }e^{it^{2}}dt.$ Rearranging complex and
complex conjugate terms, we obtain the full set of real modes of 2-D
Helmholtz equation: 
\begin{eqnarray}
&&\sin (ax\pm by),\;\cos (ax\pm by),  \TCItag{(4.8)} \\
&&\int\limits_{\sqrt{kr+ax\pm by}}^{\infty }\sin (ax\pm by-t^{2})dt\;\ \ 
\text{and }\int\limits_{\sqrt{kr+ax\pm by}}^{\infty }\cos (ax\pm by-t^{2})dt%
\mbox{.}  \nonumber
\end{eqnarray}%
One may speculate, that a fractional Laplacian operator:

\[
\Delta _{x,y}^{1/4}=_{0}{\cal L}_{x,y}^{1/4} 
\]%
has fractional eigenfunctions represented by two last terms of \ (4.8) and
the operator is related to half-plane $(x,y)$ and exhibits half-plan edge
waves. Fractional Laplacian operations may concern 2-D surfaces,\ like
circle, ring, holes in plane, etc., and they are concerned with respective
edge waves.

In the case of the 3-D Helmholtz equation and for the wave:

\[
e^{-iax-iby-icz}\int_{u(x,y,z)}^{\infty }e^{it^{2}}dt, 
\]%
we derived the following characteristic equations:

\begin{eqnarray}
u[(\partial _{x}u)^{2}+(\partial _{y}u)^{2}+(\partial _{z}u)^{2}]
&=&a\partial _{x}u+b\partial _{y}u+c\partial _{z}u,  \TCItag{(4.9)} \\
\Delta u(x,y,z) &=&0\mbox{.}  \nonumber
\end{eqnarray}

There is no known solutions to these equations. The well known solution, in
the diffraction theory, the 3-D case:

\[
e^{-iax-iby-icz}\int_{u(x,y)}^{\infty }e^{it^{2}}dt, 
\]%
where $a=k\cos \alpha \cos \beta ,\;b=\sin \alpha \cos \beta ,\;c=k\sin
\beta $ and $r=\sqrt{x^{2}+y^{2}}$, leads to $u(x,y)=\sqrt{kr\cos \beta
+ax+by\text{,}}$ which satisfies the 2-D characteristic equations and it is,
in principle, reduction of a 3-D problem by variable separation to the 2-D
problem. We note, that equations (4.9) are also obeyed.

\section{Fractional Solutions for Paraxial Propagation{{{{%
{\protect\normalsize {\bf \protect\setcounter{equation}{0}}}}}}}}

Transition from the rigorous wave theory based on a 3-D Helmholtz equation:

\begin{equation}
\nabla ^{2}U(x,yz)+k^{2}\varepsilon (x,y,z)U(x,y,z)=0\mbox{,}  \tag{(5.1)}
\end{equation}

where $\varepsilon \;$is a slowly changing dielectric permeability, to the
transversal diffusion approximation leads to the change of the kind of the
differential equations and to a new formulation of the boundary value
problem. In the contrast to the elliptic Helmholtz equation, the Schr\"{o}%
dinger type parabolic wave equations (paraxial-wave equation) describes the
evolution of the wave amplitude in process of almost unidirectional
propagation along the optical axis. Physically, it means neglecting the
backward reflections from the interfaces and an inaccurate description of
the waves diffracted into large off-axis angles. This approximate approach
finds a wide spectrum of applications in radio wave propagation, underwater
acoustics, linear laser beam propagation and X-ray imaging optics, see [13],
[14], [15], [16] and [17]. A variety of its modification has been used in
the diffraction theory, nonlinear optics and plasmas, see [18], [19], [20],
[21], [22] and [23].

Substituting the following form of expected solutions:

\[
U(x,y,z)=u(x,y,z)\exp (i{\bf k\ast r}) 
\]%
where $k=(p,q,\gamma ),\;\;|{\bf k}|=k,\;\gamma ^{2}=k^{2}-(p^{2}+q^{2}),\;%
{\bf k\ast r=}px+qy+\gamma z{\bf ,\;\sin \beta =}\frac{\sqrt{p^{2}+q^{2}}}{k}%
,$ to equation (5.1) and assuming $\varepsilon =1+\alpha (x,y,z),\;|\alpha
|\ll 1$ , sin$\beta \thickapprox 0\;,\;\partial _{z,z}u\thickapprox 0,\;$we
derive:

\begin{equation}
2ik\partial _{z}u+\Delta _{x,y}u+k^{2}\alpha (x,y,z)u=0\mbox{.}  \tag{(5.2)}
\end{equation}%
For the parabolic equation, the Cauchy problem with the given initial
distribution $\ u(x,y,0)=u_{0}(x,y)$ (named also one-point boundary value)\
is correctly posed if some radiation condition is added excluding spurious
waves coming from infinity.

Considering the two cases; (2+1)D and (1+1)D, we neglect the inhomogeneity
of the dielectric permeability $\alpha \thickapprox 0,$ and write the two
following equations:

\begin{equation}
2ik\partial _{z}u+\Delta _{x,y}u=0\mbox{.}  \tag{(5.3)}
\end{equation}

\begin{equation}
2ik\partial _{z}u+\partial _{x,x}u=0\mbox{.}  \tag{(5.4)}
\end{equation}%
Next, we factorize the equation(5.4) to obtain:

\[
_{0}D_{x}^{2}+2ik\;_{0}D_{x}^{1}=[_{0}D_{x}^{1}+\sqrt{k}%
(1-i)_{0}D_{z}^{1/2}]\ast \lbrack _{0}D_{x}^{1}-\sqrt{k}%
(1-i)_{0}D_{z}^{1/2}], 
\]%
where the fractional derivative, according to Riemann-Liouville definition,
takes the form

\[
_{0}D_{z}^{1/2}u(x,z)=\frac{1}{\sqrt{\pi }}\partial
_{z}\int\limits_{0}^{z}u(x,\xi )\frac{d\xi }{\sqrt{z-\xi }} 
\]%
and the formula is used to describe nonlocal surface admittance, see[17].

Applying the Laplace transformation with respect to the variable $z$ to the
following equation:

\begin{equation}
\lbrack _{0}D_{x}^{1}\pm \sqrt{k}(1-i)_{0}D_{z}^{1/2}]u(x,z)=0\mbox{,} 
\tag{(5.5)}
\end{equation}%
we obtain:

\begin{equation}
\lbrack _{0}D_{x}^{1}\pm \sqrt{ks}(1-i)_{0}D_{z}^{1/2}]u(x,s)=F(x)\mbox{.} 
\tag{(5.6)}
\end{equation}%
and neglecting the initial condition \ $F(x)\;$at$\;\;z=0,$ we find:

\[
u(x,z)=\frac{\sqrt{k}x}{\sqrt{2\pi iz^{3}}}\exp (\frac{ikx^{2}}{2z}). 
\]%
We now factorize (5.6) again and determine the following solution:

\begin{equation}
u(x,y)=\frac{1}{k}\sqrt{\frac{i}{\pi }}\int\limits_{x\sqrt{\frac{k}{2z}}%
}^{\infty }\exp (it^{2})dt\mbox{.}  \tag{(5.7)}
\end{equation}%
It is easy to check that (5.7) satisfies (5.4) and its first as well as
higher derivatives with respect to $x$ also\ satisfy (5.4). Therefore, we
call it the mother of solutions. By variable separation, we can write the
mother of solutions for (2+1)D equation (5.3) in the form:%
\begin{equation}
u(x,y,z)=\frac{i}{\pi }(\int\limits_{x\sqrt{\frac{k}{z}}}^{\infty }\exp
(it^{2})dt)(\int\limits_{y\sqrt{\frac{k}{z}}}^{\infty }\exp (it^{2})dt)%
\mbox{.}  \tag{(5.8)}
\end{equation}%
which is related to fractional Laplacian with respect to a certain region of
plane $(x,y)$. It will cause no confusion if we use the same notation $u$ to
designate different solutions of (5.3) and (5.4). The derivatives $\partial
_{x}^{m}\partial _{y}^{n}$ of (5.8), where $m$ and $n$ are natural numbers,
satisfy the PWE and may be related to higher order Gaussian-Hermite optical
beams.

Let us calculate the derivative $\partial _{x}\partial _{y}$ of (5.8) to
derive classical paraxial Green function:

\[
G(x,y,z)=\frac{ik}{2\pi z}\exp (\frac{ik(x^{2}+y^{2})}{2z})\mbox{.} 
\]%
It is well known that, in the problem of diffraction by plane screens (e.g.
a thin zone plate), the parabolic approximation is equivalent to the
simplified Fresnel-Kirchoff diffraction theory. In fact, any solution of
(5.3) can be expressed for $z>0$ in terms of its one-point boundary value
(initial value) over an aperture. Assuming the one-point boundary condition
over the plane $(x,y)$:

\[
u_{0}(x,y)=\exp (-\frac{x^{2}+y^{2}}{w_{0}^{2}})\mbox{,} 
\]%
where $w_{0}$ is the beam radius at $z=0$ \ and by use of the convolution
integral:

\begin{equation}
U(x,y,z)=\int\limits_{-\infty }^{\infty }(\int\limits_{-\infty }^{\infty
}u_{0}(x^{`},y^{`})G(x-x^{`},y-y^{`},z)dx^{`})dy^{`}\mbox{,}  \tag{(5.9)}
\end{equation}%
we derive:

\begin{equation}
U(x,y,z)=\frac{k}{2\pi i(\frac{2}{w_{0}^{2}}-\frac{ik}{z})}\exp (-\frac{%
k(x^{2}+y^{2})}{kw_{0}^{2}+2iz})\mbox{,}  \tag{(5.10)}
\end{equation}%
and by simple algebraic manipulation, we can write the classical form of the
laser beam, see [24] and [25]:

\begin{equation}
U(x,y,z)=\frac{w_{0}}{w(z)}\exp (-\frac{r^{2}}{w(z)^{2}})\exp (i\tan ^{-1}+%
\frac{ikr^{2}}{2R(z)})\mbox{,}  \tag{(5.11)}
\end{equation}%
where $r^{2}=x^{2}+y^{2},\,\;\;w(z)=w_{0}\sqrt{1+(z/z_{R})^{2}}\;\;$is the
beam radius,\ $z_{R}=\pi w_{0}^{2}/\lambda =kw_{0}^{2}/2\;$is the Rayleigh
length and $R(z)=z+z_{R}^{2}/z$ \ is the Rayleigh curvature. Higher order
modes can be derived by calculation of the following derivatives:

\[
(-1)^{m}(-1)^{n}\partial _{x}^{m}\partial _{y}^{n}U(x,y,z)\mbox{.} 
\]%
By inspection or simple reasoning one can see that they satisfy (5.3). In
view of the convolution properties of (5.9), we can differentiate functions $%
u_{0}(x,y)\;$or $G(x,y,z)\;$to obtain the same result

By\ a symmetry of (5.3), we find that the equation can be reduced to (1+1)D
equation (5.4) substituting $\xi =x+y.\;$In virtue of (5.7), we derive the
next mother solution:

\begin{equation}
u(x,y,z)=\frac{k}{\pi }\int\limits_{v(x,y,z)}^{\infty }\exp (it^{2})dt%
\mbox{,}  \tag{(5.12)}
\end{equation}%
where $v(x,y,z)=\frac{x+y}{2}\sqrt{\frac{k}{z}}.$ The solution (5.8) is
related to fractional Laplacian with the strip symmetry for $\left|
x+y\right| <const.\;$Higher order modes derived by the following
differentiation $\partial _{x}\;$and $\partial _{x,x}=\partial
_{x,y}=\partial _{y,y}\;$are as follows:

\[
\frac{k}{2\pi }\sqrt{\frac{k}{z}}\exp (\frac{ik(x+y)^{2}}{4z})\;\text{and \ }%
i\frac{k^{2}}{4\pi z}\sqrt{\frac{k}{z}(x+y)}\exp (\frac{ik(x+y)^{2}}{4z})%
\mbox{,} 
\]%
and solutions to one-point boundary value problems, e.g. $%
exp[(x+y)^{2}/w_{0}^{2}]$, can be derived by use of Fresnel-Kirchoff
integral (5.9). By symmetry consideration, see [23], it seems that the
exhibited mother solutions for PWE exhaust all possibilities.

We now give an example of the Fresnel solution to the following
nonhomogeneous PWE:

\begin{equation}
(2ik\partial _{z}+\Delta _{x,y})V(x,y.z)=-\frac{a}{\sqrt{x^{2}+y^{2}}}\sqrt{%
\frac{k}{2z}}\exp (ik\frac{x^{2}+y^{2}}{2z})\mbox{,}  \tag{(5.13)}
\end{equation}%
where a is a constant.

The solution of (5.13) for the homogeneous initial condition; $U_{0}(x,y)=0$%
, takes the very interesting form of the Fresnel beam:

\begin{equation}
U(x,y,z)=a\int\limits_{v(x,y,z)}^{\infty }e^{it^{2}}dt\mbox{,}  \tag{(5.14)}
\end{equation}%
where $v(x,y,z)=\sqrt{\frac{k}{2z}}\sqrt{x^{2}+y^{2}}.$

The real and imaginary part of (5.14) as well as the field intensity $\left|
U(x,y,z)\right| $ are depicted in Fig. 1-3.

\begin{center}
\FRAME{ihFU}{2.258in}{1.8377in}{0in}{\Qcb{Fig.1 3-D diagrams illustrating
the real part of the Fresnel beam $\func{Re}U=a\protect\int%
\limits_{v(x,y,z)}^{\infty }\cos (t^{2})dt$, where $v(x,y,z)=\protect\sqrt{%
k/2z}\protect\sqrt{x^{2}+y^{2}},$ $k=2,\;z=2\;$and $a=1$.\ }}{\Qlb{Fig.1}}{%
fig1.gif}{\special{language "Scientific Word";type
"GRAPHIC";maintain-aspect-ratio TRUE;display "PICT";valid_file "F";width
2.258in;height 1.8377in;depth 0in;original-width 4.0101in;original-height
3.2603in;cropleft "0.02922";croptop "1.015989";cropright
"1.02922";cropbottom "0.015989";filename 'Fig1.gif';file-properties "XNPEU";}%
}
\end{center}

\FRAME{ihFU}{2.5175in}{2.0435in}{0in}{\Qcb{Fig. 2. \ 3-D diagrams
illustrating the imaginary part of the Fresnel beam $\func{Im}U=a\protect\int%
\limits_{v(x,y,z)}^{\infty }\sin (t^{2})dt$, where $v(x,y,z)=\protect\sqrt{%
k/2z}\protect\sqrt{x^{2}+y^{2}},$ $k=2,\;z=2\;$and $a=1$.}}{\Qlb{Fig.2}}{%
fig2.gif}{\special{language "Scientific Word";type
"GRAPHIC";maintain-aspect-ratio TRUE;display "PICT";valid_file "F";width
2.5175in;height 2.0435in;depth 0in;original-width 4.3439in;original-height
3.5206in;cropleft "0";croptop "1";cropright "1";cropbottom "0";filename
'Fig2.gif';file-properties "XNPEU";}}

\FRAME{ihFU}{2.1153in}{1.7193in}{0in}{\Qcb{Fig.3.\ 3-D diagrams illustrating
the field intencity $\left| U\right| =\protect\sqrt{(\func{Re}U)^{2}+(\func{%
Im}U)^{2}}$ of the Fresnel beam.}}{\Qlb{Fig.3}}{fig3.gif}{\special{language
"Scientific Word";type "GRAPHIC";maintain-aspect-ratio TRUE;display
"PICT";valid_file "F";width 2.1153in;height 1.7193in;depth
0in;original-width 4.6873in;original-height 3.8017in;cropleft "0";croptop
"1";cropright "1";cropbottom "0";filename 'Fig3.gif';file-properties
"XNPEU";}}

\section{Comments and Conclusions{{{{{\protect\normalsize {\bf \  \protect
\setcounter{equation}{0}}}}}}}}

In principle, the number of eigenfunctions of the differential Laplace
operator can be arbitrary. According to factorization and a choice of $%
D^{1/n}=a,$ the candidate number is $n$ for the Laplacian. In the case of
1-D and 2-D, we derive 4 semi-differential modes. The 2-D Laplacian $\Delta
_{x,y}$ possesses two additional eigenfunctions of the form;

\[
\int\limits_{u(x,y)}^{\infty }\sin (ax\pm by-t^{2})dt\;\ \ \text{and }%
\int\limits_{u(x,y)}^{\infty }\cos (ax\pm by-t^{2})dt\mbox{,} 
\]%
where the lower limit of Fresnel integral $u(x,y)\;$is to satisfy the set of
characteristic equations, see (4.5), for the half plane $\ x,y\in (0,\infty
).$ It is important to note, that the modes are related to a half-plane,
like fractional derivatives are related to an interval. We recall, as an
example, that the fractional derivative of order $q$ of the exponential
function $e^{\lambda x}$ related to a half-axis is a Mittag-Leffler function
but for the whole axis it is the same function multiplied by $\lambda $ to
power $q$; $\lambda ^{q}e^{\lambda x}$. We do withhold, throughout the
paper, \ from defining fractional Laplacian operator although we determine
the eigenfunction of the operator related to half-plane, strip and circle.
There is an expectation, that the right definition must be based on an
integral of the Fresnel-Kirchoff type and relation to the $n$-dimensional
region, where $n$ is a number of independent variables. One may speculate,
on the ground of Riemann-Liouville definition, that it may be a two-folded
convolution integral with respect to $x$ and $y$ such that a Laplace
transform with respect to the variables gives an anticipated results like in
the case of single variable functions. Also, one may hope, that the Gr\"{u}%
nwald-Letnikov definition may be extended for the fractional operators. A
separate paper will be devoted to existence, uniqueness and definitions of
fractional Laplacian operators. Here considered operators come out from
factorizations and lead to results, which satisfy the classical Laplacian.
It is not a necessary condition for the fractional operators.

We also mention the 3-D Helmholtz equation but for that problem there is no
expected solution. We do not know the 3-D region, like in the case of
half-plane diffraction, for which there is an exact solution
(eigenfunctions). The solution derived by variable separation (a method
reducing the 3-D problem to two dimensions) is not applicable to the our
requirement.

Consideration of Leontovich-Fock equation is justified not only by its
numerous applications but to show that the equation is factorable and the
fractional equations lead to the mother of solutions. The mother is a
function of the Fresnel integral satisfying PWE, vanishing at $z=0$ and her
derivatives $\partial _{x}^{m}\partial _{x}^{n}$ also satisfy PWE. The
notion is not trivial as there is expectation that the mother solutions and
their derivatives generate all possible symmetries of on-axis and off-axis
beams in the case of homogeneous PWE. The last solution (5.14) of
nonhomogeneous equation (5.13) is an illustration of paraxial beam of the
Fresnel type.

\bigskip

{\Large References}

\begin{enumerate}
\item {\normalsize {\sc K.B. Oldham }and {\sc J. Spanier}, {\sl The} {\sl %
Fractional} {\sl Calculus, }Academic Press, New York-London, 1974. }

\item {\normalsize {\sc S.G. Samko, A.A. Kilbas }and {\sc O.I. Maritchev}, 
{\sl Integrals and Derivatives of \ the Fractional Order and Some of their
Applications,} [in Russian], Nauka i Tekhnika, Minsk, 1987.}

\item {\normalsize {\sc K.S. Miller} and {\sc B. Ross}, {\sl An Introduction
to the Fractional Calculus and Fractional Differential Equations, } John
Wiley \& Sons Inc., New York, 1993. }

\item {\normalsize {\sc Igor Podlubny}, An {\sl Introduction to Fractional
Derivatives, Fractional Differential Equations, to Methods of their Solution
and some of their Applications}, Academic Press, New York-London, 1999. }

\item {\normalsize {\sc R. Hilfer}, Editor, {\sl Applications of Fractional
Calculus in Physics}, Word Scientific Publishing Co., New Jersey,
London,elmh6b Hong Kong, 2000. }

\item {\sc J.Spanier and K.B. Oldham, }{\normalsize {\sl An Atlas of
Functions, }}Springer-Verlag, Berlin- Tokyo, 1987.

\item {\normalsize {\sc F. Mainardi}, {\sl On the Initial Value Problem for
the Fractional Diffusion-wave Equations, in; S. Rionero and T. Ruggeri
(eds.); Waves and Stability in Continuous Media, \ }}Word Scientific,
Singapore, 246-251, 1994.

\item {\normalsize {\sc F. Mainardi}, {\sl The Fundamental Solutions for the
Fractional Diffusion-wave Equation, }}Appl. Math. Lett., vol. 9, no. 6,
23-28, 1996.

\item {\sc A}{\normalsize {\sc . Carpintery }}and{\normalsize {\sc \ F.
Mainardi (eds).}, {\sl Fractals and Fractional Calculus in Continuum
Mechanics, }}Springer Verlag, Vienna-New York, 1997.

\item {\normalsize {\sc F. Mainardi}, {\sl Fractional Relaxation-oscillation
and Fractional Diffusion-wave Phenomena. }}Chaos, Solitons and Fractals,
vol. 7, 1461-1477, 1996.

\item {\sc M. Seredynska {\normalsize {\sl and }}}{\normalsize {\sc A.
Hanyga, }{\sl Nonlinear Hamiltonian Equations with Fractional Damping, }}J.
Math. Phys. 41, 2135-2156, 2000.

\item M. BORN and E. WOLF,{\it ''Principles of Optics''}, Pergamon Press,
1964.

\item {\sc V. A. Fock, }{\normalsize {\sl Electromagnetic Diffraction and
Propagation Problems, }}Pergamon Press, Oxford, 1965.

\item {\sc E. D. Tappert, }{\normalsize {\sl The Parabolic Approximation
Method, Lectures Notes in Physics, 70, in: Wave Propagation and Underwater
Acoustics, }}eds. by{\normalsize {\sl \ }}{\sc J. B. Keller and J. S.
Papadakis, }Springer, New York, 224-287, 1977.

\item {\sc L. A. Vainstein, }{\normalsize {\sl Open Resonators and Open
Waveguides, (in Russian)\ }}Soviet Radio, Moscow. 1966.

\item {\sc S. W. Marcus, }{\normalsize {\sl A Generalized Impedance Method
for Application of the Parabolic Approximation to Underwater Acoustics, }}J.
Acoust. Soc. Am. {\bf 90}, no.1 391-398, 1991.

\item {\sc A. V. Vinogradov, A. V. Popov, Yu. V. Kopylov and A. N.
Kurokhtin, }{\normalsize {\sl Numerical Simulation of X-ray Diffractive
Optics, }}A\&B Publishing House Moscow 1999.

\item {\sc G. D. Malyuzhinets, }{\normalsize {\sl Progress in understanding
diffraction phenomena, (in Russian), }}Soviet Physics (Uspekhi), {\bf 69},
no.2, 312-334, 1959.

\item {\sc A. V. Popov, }{\normalsize {\sl Solution of the parabolic
equation of diffraction theory by finite difference method, (in Russian), }}%
J. Comp. Math. and Math. Phys., 8, no.5, 1140- 1143, 1968.

\item {\sc V. M. Babich and V. S. Buldyrev,}{\sl \ }{\normalsize {\sl %
Short-wavelength diffraction theory, (Asymptotic methods), }}Springer, New
York, 1991.

\item {\sc V. E. Zakharov, A. B. Shabat, }{\normalsize {\sl Exact theory of
two-dimensional self-focusing and unidimensional self-modulation of waves in
nonlinear media, (in Russian), }}JETP, {\bf 61, }no.1(7), 118-1134, 1971.

\item {\sc W. Nasalski, }{\normalsize {\sl Beam switching at planar photonic
structures,}} Opto-Electronics Review, {\bf 9}(3),280-286, 2001.

\item {\sc Z. J. Zawistowski and A. J. Turski, }{\normalsize {\sl Symmetries
of nonlocal NLS equation for Langmuir waves in Vlasov plasmas,\ }}J.Tech.
Phys. {\bf 39}, 2,297-314, 1998.

\item {\sc A \ E. Siegman, }{\normalsize {\sl Lasers,\ }}University Science
Books, 1986.

\item {\sc J.\ T. Verdeyen, }{\normalsize {\sl Lasers Electronics,\ }}%
University of Illinois, 1993.
\end{enumerate}

{\normalsize \ \ \noindent {\small {\bf POLISH ACADEMY OF SCIENCES}}\newline
{\small {\bf INSTITUTE OF FUNDAMENTAL TECHNOLOGICAL RESEARCH}}}
\begin{verbatim}
E-mail: aturski@ippt.gov.pl
\end{verbatim}

\end{document}